# Stacking and electric field effects on the electronic properties of the layered GaN

Dongwei Xu[1], Haiying He[1*], Ravindra Pandey[1*], and Shashi P. Karna[2]

[1]*Department of Physics, Michigan Technological University, Houghton, Michigan 49931, USA*

[2]*US Army Research Laboratory, Weapons and Materials Research Directorate, ATTN: RDRL-WM, Aberdeen Proving Ground, MD 21005-5069, U.S.A.*

(February 20, 2013)

*Corresponding Authors:
Haiying He: hhe@mtu.edu
Ravindra Pandey: pandey@mtu.edu

## Abstract

Stability and electronic properties of atomic layers of GaN are investigated in the framework of the van der Waals-density functional theory. We find that the ground state of the layered GaN is a planar graphene-like configuration rather than a buckled bulk-like configuration. Application of an external perpendicular electric field to the layered GaN induces distinct stacking-dependent features of the tunability of the band gap; the band gap of the monolayer does not change whereas that of the trilayer GaN is significantly reduced for the applied field of ±0.4 V/ Å.  It is suggested that such a stacking-dependent tunability of the band gap in the presence of an applied field may lead to novel applications of the devices based on the layered GaN.

# 1. Introduction

Gallium nitride (GaN) is a semiconductor of great interest due to its unique optoelectronic properties, high mechanical stability and good thermal conductivity with applications ranging from laser diodes, high-electron-mobility transistors, to solar cells.[1-3] A large and direct band gap[4] (≈3.4 eV) of GaN also makes it to be an excellent host material for light-emitting devices (LED) that operate in the blue and ultraviolet region.[5] Note that the traditional high-brightness LEDs consisted of a thin film of GaN deposited on sapphire. A recent break-through was made to use silicon as a substrate instead of sapphire for growing GaN LEDs, thus making them to be economically competitive for the large-scale day to day applications.[6]

As a key to the optoelectronic applications, an ability to continuously control the band gap in a wide spectrum is highly desirable. For the novel 2-d planar graphite-like structure, such as graphene, multilayer BN or BN-graphene hybrid structures, the band gap can be tuned by an applied electric field.[7-11] It is now expected that recent advances in synthesis and fabrication of GaN with nanometer-scale thickness[12-15] can facilitate a control over the tunability of its band gap.

Considering that a 2-d atomic layer represents the extreme case of an ultrathin film of GaN, we take a bottom-up approach to investigate stability, structural and electronic properties of GaN built from a monolayer to bilayer to trilayer. Since the reduction of the dimension may drastically change the physical properties of the material, we would also like to address the following questions: (i) How is the structure of a GaN multilayer (composed of a few hexagonal atomic layers) different from the bulk material? (ii) What is the change in their electronic structure and how does it depend on the stacking? (iii) Can the band gap of the layered GaN be tuned by applying an electric field and to what extent? It is to be noted here that tunability of a monolayer of GaN under external electric field has been considered previously.[16, 17] We believe that a detailed understanding of these low-dimensional building blocks of GaN may help us to develop nanoscale optoelectronic devices utilizing the advanced modern fabrication techniques.

The computational details and validation of our method are given in Sec. 2. In Sec.

3.1, we present the results of structural evolution of GaN atomic layers from a single layer up to three layers. The stacking alternatives of the layered GaN will be fully explored. The electronic properties and their dependence on the stacking sequence are presented in Sec. 3.2. In Sec. 3.3, the response of GaN multilayer under an external perpendicular electric field is reported. A brief summary of the results is given in Sec. 4.

**2. Method**

The projector augmented plane-wave (PAW) method as implemented in Vienna *ab initio* simulation package (VASP) is employed.[18, 19] The generalized gradient approximation (GGA) of Perdew, Burke and Ernzerhof (PBE) is adopted for the exchange-correlation functional[20] to the density functional theory (DFT), together with the van der Waals (vdW) interactions described via a pair-wise force field in the DFT-D2 method of Grimme.[21] The 2s and 2p electrons of nitrogen, 3d, 4s and 4p electrons of Ga are considered explicitly as the valance electrons for the electronic structure calculations.

A minimum vacuum distance of 12 Å between neighboring images is used in the supercell employed. The cut-off energy for the plane-wave basis set is set to 520 eV. The energy tolerance is $10^{-6}$ eV in the iterative solution of the Kohn-Sham equations. The structure and atoms are relaxed until the force on each atom is less than 0.01eV/Å. The Monkhorst-pack mesh of (15x15x1) is selected in the Brillouin-zone integrations for the structural optimization and density of state (DOS) calculations. The calculated equilibrium configurations with or without external electric field are fully relaxed. It should be pointed out that the GGA-DFT method has been shown to provide a reasonably good description of the physics and chemistry of GaN systems although it underestimates the band gap of the semiconducting and ionic materials.[17]

In order to validate our modeling elements, the calculated structural and electronic properties of the most common wurtzite structure of the bulk GaN are compared with the experimental and previously reported theoretical studies. The calculated lattice constants, a=3.204 Å (3.215 Å), c=5.240 Å (5.230 Å) at the vdW-DFT (GGA-DFT) level of theory are in good agreement with the corresponding experimental values of a=3.185 Å, c= 5.185 Å.[4, 22] The vdW-DFT band gap is 1.74 eV, which is in line with the previous DFT calculation,[23] but lower than the experimental value[4] of 3.4 eV. Standard DFT functional forms such as GGA, in general, underestimate the band gap of semiconductors,[24] as our calculation also indicates here. Despite the underestimation of the absolute value of band gaps, however, DFT calculations still provide a computationally efficient way to successfully capture the qualitative changes in band structures of semiconducting materials.

## 3. Results and Discussion
### 3.1 Structural properties

A single layer of GaN perpendicular to the (001) direction is cleaved from the wurtzite GaN, and fully relaxed during the geometry optimization. Unlike the bulk wurtzite structure where each atom is four-fold coordinated, the monolayer GaN takes a planar graphene-like structure with three-fold coordination. The calculated Ga-N bond length is 1.85 Å, which is smaller than the bulk value of 1.96 Å. Note the bond lengths calculated at either the GGA-DFT or vdW-DFT level of theory are the same, suggesting the presence of the predominant ionic bonding in a single atomic layer as the case with the bulk GaN. These results are comparable with our previous results[17] obtained using the SIESTA code at the GGA-DFT level of theory, where the bond length is reported to be 1.91 Å for the monolayer GaN.

By bringing monolayers together, one can form a bilayer. There exists five possible stacking sequences for such a bilayer system, i.e., AA, AA', AB (NN), AB (GaGa), AB (NGa), as shown in Fig. 1. Similar to a hexagonal monolayer of BN, the monolayer of GaN can be divided into two sublattices; Ga atoms occupy one sublattice while N atoms occupy the other. In a bilayer, we label the atoms of the bottom and top layers as X1 (X=N or Ga) and X2 (X=N or Ga), respectively. For the AA' (hexagonal) stacking, the sublattice N2 (Ga2) is located directly on top of Ga1 (N1). We have also considered the AA stacking, where the sublattice N2 (Ga2) is located directly on top of N1 (Ga1). For the AB (Bernal-type) stacking, only one sublattice of the top layer is located on the top of a sublattice of the bottom layer and the other sublattice is on top of the center of the hexagon of the bottom layer. If the atoms on top of each other are N2 and N1, it is designated as AB (NN) stacking. If they are Ga2 and Ga1, it is labeled as AB (GaGa) stacking. If they are of different species, it is designated as AB (GaN) that is essentially the same as AB (NGa) (Fig. 1).

By adding one more layer to a bilayer, a trilayer is formed for which eight stacking sequences are considered as shown in Fig. 2. The GaN trilayers are constructed in such a way that the stacking order of the neighboring two layers follows one of the AA, AA' and AB (NGa) stackings of the bilayers. For instance, AA' stacking of the lower two layers and AA stacking of the top two layers form the AA'A stacked trilayer. Similarly, AA+AA form AAA, AA+AB form AAB(NNGa) or AAB (GaGaN) depending on the relative shift of the top layer. A'A+AA can form A'AA, and AB+AB can form three different stacked configurations for the trilayer GaN.

The optimized bond lengths, interlayer distances, cohesive energies ($E_{coh}$) of the bi- and trilayer configurations of GaN are listed in Table 1. The cohesive energy is defined as

$$E_{coh} = \left(\sum_i^{N_{pair}}(E_N + E_{Ga}) - E_{total}\right)/N_{pair}, \qquad (2)$$

where $E_N$ and $E_{Ga}$ are the energy of the isolated atom N and Ga, and $N_{pair}$ is the number of N-Ga pairs in calculating the total energy of the system. This intrinsic parameter reflects the stability of the multilayer systems. In the following, we will primarily use the vdW-DFT results for discussion. A comparison of the vdW-DFT results with the GGA-DFT results will be given at the end of this section.

For the bilayer GaN, we find that the hexagonal AA' stacking is the most energetically favorable ($E_{coh}$=8.68 eV/Ga-N pair), followed by the Bernal-type AB (GaN) stacking ($E_{coh}$=8.58 eV/Ga-N pair). The AA, AB (NN) and AB (GaGa) stackings are found to be relatively less stable (Table 1). The order of stability is also reflected in the calculated interlayer binding energy (0.58 eV and 0.41 eV for AA' and AB (GaN) stackings, respectively) and the interlayer separation (2.47 Å and 2.94 Å for AA' and AB (GaN) stackings, respectively). The smaller the interlayer separation, the higher the binding energy is. The interlayer binding energy with reference to the constituent layers offers a direct gain in the overall cohesive energy of the system. The magnitude of the interlayer binding energy for different stackings can be traced back to the contributions of the electrostatic and vdW interactions. In the semi-ionic GaN lattice, the cation (gallium) tends to attract an anion (nitrogen), and repel another cation. Thus, in the AA' stacking, each Ga (N) sublattice of one layer tends to attract the adjacent N (Ga) sublattice of the other layer, resulting in a small interlayer spacing and a higher interlayer binding energy. In the AA stacking, each Ga (N) sublattice of one layer tends to repel the adjacent Ga (N) sublattice of the other layer, resulting in a large interlayer spacing and barely any interlayer binding energy. Very strikingly, the effect of vdW interactions is not negligible in this case and the interlayer spacing reduces from 4.66 to 3.63 Å after including the vdW terms in DFT. Nonetheless, the intralayer bond length gets stretched slightly with respect to that of a monolayer (≈1.85 Å) at the vdW-DFT level of theory.

For GaN trilayers, AA'A is found to be the most stable stacking configuration, and is followed by A'AA, ABA (or ABC), AAB and AAA stackings. The stability of the AA'A stacked trilayer is likely to be due to the alternating cation-anion arrangements between the layers (Fig. 2). The interlayer binding energy in this case is 1.20 eV, the highest of all. As a result, the AA'A stacking has the smallest interlayer distance and largest intralayer bond length amongst the trilayer configuration considered. Note that there exist two values for the interlayer spacing for A'AA because of the different stacking order between the 1st and the 2nd layer, and between the 2nd and the 3rd layer (Table 1).

Our results therefore find the most stable stacking sequence of GaN atomic layers to be AA'AA'.... The calculated cohesive energy slowly increases with the number of layers; monolayer (8.38 eV/Ga-N pair), the AA' bilayer (8.68 eV/Ga-N pair), and the AA'A trilayer (8.78 eV/Ga-N pair) indicating the increasing stability of the multilayer system with increasing the number of layers. Thus, the hexagonal stacked configuration is predicted to be the most stable configuration for the layered GaN. This is in contrast to the case of the bilayers of graphene[25] and BN[26] where the Bernal stacking forms the ground state. On the other hand, both Bernal and rhombohedral stackings are stable for the trilayer grapheme.[27-29] We attribute the difference in graphene (BN) and GaN to their ionicity. The stronger ionic nature of Ga and N atoms renders the AA' and AA'A stackings to be the more stable configurations for the layered GaN.

It is interesting to see a similarity in the stacking sequence shared by the layered GaN and the bulk GaN, though the layers are buckled in the bulk GaN. Nonetheless, it is worth noting that the planar structure is always the preferred structure for both the bilayer and trilayer configurations of GaN.  There may exist a small buckling in which a maximum variation in the interlayer spacing is less than 1%.  The predicted preference of planar configuration is further verified by taking a buckled bilayer directly cut from the bulk wurtzite GaN as the initial guess configuration for the unconstrained geometry optimization. The buckled bilayer relaxes to a nearly planar configuration in which the buckling is reduced from 0.67 Å to about 0.067 Å. This is different from what has been reported for the GaN nanoribbons[30] where the bilayer nanoribbons cleaved from the bulk GaN are predicted to be more stable as compared to the planar nanoribbons. We believe that the difference comes from the periodic 2-d sheet (our model) and 2-d nanoribbon with edges.[30] The existence of edges may be the cause of the buckling in nanoribbons. We note that the preference of the planar configuration is consistent with other theoretical studies on the GaN sheet.[31] Furthermore, research on the structural evolution of two-dimensional films have shown that the system prefers a planar configuration when the number of layers is less than six.[32]

Finally, our calculations also demonstrate the importance of the vdW terms for the layered structures, even in a semi-ionic material, like GaN.  Table 1 gives the results at vdW-DFT and GGA-DFT level of theory showing the same order of the stacking-dependent stability for the layered GaN. A comparison of the structural properties suggests that the vdW terms do not change the intraplanar bond lengths, though it influences significantly the interplanar separations. The effect is more prominent for the cases where there is no bonding or barely any electrostatic interactions between GaN layers, such as the AA, AB (NN) and AB (GaGa) bilayers (Table 1). Inclusion of the vdW interaction terms drastically reduces the optimal interlayer distance by more than 1 Å with an increase in the interlayer binding of about 0.2 eV. Note that Marom *et al*. have also pointed out the crucial role of the vdW forces in anchoring the layers at a fixed distance for the BN layered structures.[33]

**3.2 Electronic structure**

We now consider the configurations obtained at the vdW-DFT level of theory for calculations of electronic properties of the layered GaN. The calculated band structures of monolayer, bilayer and trilayer configurations are shown in Fig. 3 (solid lines).

The monolayer has an indirect band gap of 2.17 eV. The conduction band minimum (CBM) is located at Γ whereas the valence band maximum (VBM) is located at K. A higher value of the band gap for the monolayer relative to that of the bulk wurtzite GaN can be attributed to the quantum confinement effect.[17] Similar to the monolayer, an indirect band gap with the values of 2.01 and 1.69 eV for the AA' and AB (GaN) stacked configurations, respectively is predicted. For the AB (GaN) bilayer, lowering of the symmetry breaks degeneracy between the top two valance bands at K. For the AA'A and ABA (NGaN) stacked trilayers, the band gap is indirect with the values of 1.85 and 1.60 eV, respectively.

A clear trend of change in electronic properties with the number of layers is noted. The bulk GaN can be deemed as an AA'-stacked (buckled) configuration with infinite number of layers. It has a Ga-N bond length of 1.97 Å and a calculated band gap of 1.74 eV. The AA'A-stacked trilayer has an intralayer Ga-N bond length of 1.89 Å and a gap of 1.85 eV. The AA'-stacked bilayer has an intralayer bond length of 1.88 Å and a gap of 2.01 eV. Finally, the monolayer has a bond length of 1.85 Å and a gap of 2.17 eV. The calculated results therefore show that the decrease in the intralayer bond length is accompanied by an increase in the band gap as the interlayer interaction is gradually removed in the layered GaN.

In order to understand the predicted variation of the band gap as a function of the number of layers, the atom-resolved projected density of states (PDOS) for monolayer, the AA' bilayer and AA'A trilayer configurations are shown in Fig. 4. In all cases, the top of the valance band is formed by the nitrogen atoms. For the AA'-stacked bilayer, the top and bottom layers have a point symmetry. As a result, we find the degeneracy between the energy spectrum of the gallium (nitrogen) atoms of the top layer and that of the gallium (nitrogen) atoms of the bottom layer. The interlayer interaction appears to increase the width of the valence band from 6.06 eV for the monolayer to 6.45 eV for the bilayer. The conduction band states associated with Ga and N atoms are also broadened reflecting the presence of the interlayer interaction in the bilayer GaN.

There exists an intimate relation between the band structure and the inter-atomic interactions in these atomic layered structures as explained by a simple and intuitive orbital interaction model (OIM) (Fig. 5) showing the shift of energy levels as a result of the interlayer interaction. The valence electrons are only considered, and the energy levels are referred to the average value of the corresponding energy spectra. Due to the different electronegativity, the energy levels (4s4p) of gallium are higher than those of nitrogen (2s2p). For simplicity, one representative level is shown for each case in the energy diagram ignoring the hybridization of gallium and nitrogen orbitals. In a monolayer, the energy level of gallium shifts up and forms the conduction band minimum (CBM), while the energy level of nitrogen shifts down and forms the valence band maximum (VBM) due to the gallium-nitrogen interaction (Fig. 5, left). For the AA' bilayer, the interlayer interaction between Ga1 (N1) and N2 (Ga2) increases the opening of the gap. However, in-plane interaction between N1 (N2) and Ga1 (Ga2) gets weaker due to increase in the Ga-N bond length from 1.85 Å (monolayer) to 1.88 Å (bilayer). At the interlayer level, on the other hand, N1 (Ga1) is symmetric to N2 (Ga2) resulting in to a degenerate eigenvalues corresponding to the same species of layers. Thus, the degenerate energy levels of Ga1 and Ga2 shifts down forming CBM while those of N1 and N2 shifts up forming the VBM resulting into a smaller band gap for the bilayer as shown in Fig 5 (middle upper).

For the AA'A stacking, degeneracy between Ga1 and Ga2 (N1 and N2) states is removed. Ga2 which interacts with both N1 and N3 goes up while Ga1 goes down, resulting in a lowering of the band gap as compared to the AA' stacked bilayer. In addition, the third layer also introduces a new degeneracy between Ga1 and Ga3 (N1 and N3) states (Fig. 5, middle lower).

### 3.3 Electronic structure under the applied electric field

We now investigate the response of the electronic structure of the GaN multilayers in the presence of an external electric field. We define the electric field from bottom to top to be positive and use the values from -0.4 V/Å to 0.4 V/Å.

Fig. 3 (dashed lines) shows the band structure of layered GaN at $E_{field}=\pm0.4$ V/Å. For the monolayer, we do not see noticeable difference in the band structure relative to that obtained at $E_{field}=0$ (solid lines). For the AA' bilayer, the responses under positive and negative electric fields are symmetric, and the band gap reduces from 2.01 eV to 1.74 eV. Note that the top and bottom GaN layers exhibit a point symmetry in AA'. This is also the case with the trilayer AA'A where the positive and negative electric field values influence in a similar way because of the symmetry; VBM moves away from Γ to K lowering the band gap from 1.85 eV ($E_{field}=0$) to 0.96 eV ($E_{field}=\pm0.4$ V/Å). For the AB (NGa) stacking, almost no changes are seen under applied field strength of 0.4 V/Å. In contrast, application of the field of ±0.4 V/Å drastically reduces the band gap from 1.60 eV ($E_{field}=0$ V/Å) to 0.33 eV for the ABA (NGaN) trilayer. As discussed above, there is less electrostatic interlayer interaction in the ABA stacking as compared to the AA'A stacking. This allows a larger polarization across the layers under the external electric field. As a result, the band gap in ABA reduces to a larger extent relative to that in AA'A.

Fig. 6 shows the calculated charge density of the layered GaN. Electrons are localized in basins around the N atoms (the red spots). A clear deficit of electron is seen around Ga atoms (the blue spots) suggesting the presence of ionic bonding considering the difference in electronegativity of Ga and N. An overlap is also seen along the vertical direction for the AA' and AA'A stackings indicating a noticeable Ga-N interlayer interaction. A smaller overlap seen for the AB (GaN) and ABA (NGaN) stackings suggests the presence of a weaker interlayer interaction in such layered GaN.

Fig. 7 shows the stacking-dependent variation of band gap for the layered GaN under a perpendicularly applied electric field. A degree of tunability of the band gap gets larger with the increase of number of layers in the configuration. For the monolayer, the AA' bilayer and the AA'A trilayer (left pane), a symmetric variation in the band gap is predicted for the positive and negative bias. For the Bernal stackings, the gap tunability shows asymmetric and inconspicuous behavior for the AB (GaN) stacking. However, variation in the band gap is symmetric and significant for the ABA (NGaN) stacking for which the gap closes at $E_{field}=\pm0.45$ V/Å by extrapolation. On the other hand, a closing of the band gap is predicted at about $E_{field}=\pm0.58$ V/Å for the AA'A trilayer.

Our simplified OIM model (Fig. 5) provides a good explanation of variation in the band gap with the applied field. Taking bilayer as an example, under a positive bias, both the valence and conduction bands of the top layer shift to a lower energy, while the bands of the bottom layer shift to a higher energy. CBM of AA' primarily consists of electronic states from Ga2, whereas VBM consists of the N1 states. With the increase of the applied field, the energy levels of Ga2 and N1 get closer to each other resulting in the lowering of the band gap. For the AA'A stacking, degeneracy of the energy levels between the bottom and top layers breaks, though the levels of the middle layer are almost unchanged. Using a similar approach as used in the bilayer case, one can derive that degree of reduction in the band gap is determined by Ga3 and N1 (Fig. 5, right). Note that our results on the bilayer GaN are consistent with the previously reported study on the bilayers of boron nitride.[7]

## 4 Summary

The stacking-dependent structural and electronic properties of the layered GaN are investigated using the vdW-DFT level of theory. The AA' and AA'A stackings are predicted to be the most stable configurations for bilayer and trilayer of GaN, respectively. A symmetric variation of the band gaps of AA' and AA'A is predicted with the applied electric field. Overall, the band gap decreases with the increase in the electric field. The simple OIM model can be used to understand the predicted variation, while a shift in the energy levels is confirmed by analysis of PDOS. It is notable that the band gap can be varied by changing the stacking orders of the layered GaN. A response to an external electric field applied perpendicular to the layer is the largest for the ABA (NGaN) trilayer. Furthermore, the calculated results predict tuning of the band gap from 1.6 to 0.33 eV with $E_{field}$ varying from 0 to ±0.4 V/Å.

It is expected that the calculated results may provide a fundamental basis for applications of the layered GaN as a candidate semiconducting material for novel optoelectronic devices. With great advances recently made in fabrication of GaN based nanostuctures, e.g. nanorods,[34, 35] nanowires,[36, 37] nanosaws[38] and nanotubes,[39] an accurate control of the stacking sequence and the thickness of these GaN layered structures may soon become possible.

**Acknowledgments**

Helpful discussions with S. Gowtham and Xiaoliang Zhong are acknowledged. The work at Michigan Technological University was performed under support by the Army Research Laboratory through Contract Number through contract number W911NF-09-2-0026.

Figure Captions:

Figure 1. (Color online) Stacking configurations considered for the bilayer of GaN. The large brown balls represent 'Ga' while the small pink balls represent 'N'.

Figure 2. (Color online) Stacking configurations considered for the trilayer of GaN. The large brown balls represent 'Ga' while the small pink balls represent 'N'.

Figure 3. (Color online) The calculated band structures of the monolayer (upper panes), the bilayers (middle panes) and the trilayers (bottom panes) of GaN with hexagonal and Bernal stackings. Solid (black), dashed (blue) and dot-dashed (red) lines represent the applied field of 0 V/Å, 0.4V/Å and -0.4V/Å, respectively. The inset shows the zoom-in near the conductance band minimum for the AB stacked GaN.

Figure 4. (Color online) The atom-resolved density of states of the monolayer, the AA'-stacked bilayer and the AA'A-stacked trilayer of GaN.

Figure 5. Schematic representation of the OIM model of the monolayer, the AA' stacked bilayer and the AA'A stacked trilayer of GaN.

Figure 6. (Color online) Charge distributions of the monolayer (upper pane), the bilayers (middle pane) and the trilayers (lower pane) of GaN with hexagonal and Bernal stackings. The 2D plane is in the $<1\bar{1}0>$ direction. The color scale is labeled in unit of $e/$Bohr$^3$.

Figure 7. (Color online) Variation in the band gap with the applied electric field for the layered GaN: monolayer in black square, AA' bilayer in red up-triangle, AB (GaN) bilayer in red circle, AA'A trilayer in green diamond, and ABA (NGaN) trilayer in green down-triangle.

Table 1 Intraplanar bond length $R_{bond}$, interplanar separation $R_{interlayer}$, cohesive energy $E_{coh}$ and binding energy $E_b$ for the monolayer, bilayer and trilayer GaN after full optimization by GGA-DFT and GGA+vdW-DFT methods. In the third column showing the corresponding figures, small purple ball represents 'N' and large brown ball represents 'Ga'.

| | Stackings | Representative Figure | Method | $R_{bond}$ (Å) | $R_{interlayer}$ (Å) | $E_{coh}$ (eV/pair) |
|---|---|---|---|---|---|---|
| Monolayer | | | GGA - DFT | 1.85 | - | 8.14 |
| | | | vdW-DFT | 1.85 | - | 8.38 |
| Bilayer (hexagonal type) | AA | | GGA - DFT | 1.85 | 4.66 | 8.14 |
| | | | vdW-DFT | 1.85 | 3.63 | 8.46 |
| | AA' | | GGA - DFT | 1.89 | 2.40 | 8.14 |
| | | | vdW-DFT | 1.88 | 2.47 | 8.68 |
| Bilayer (Bernal type) | AB (GaN) | | GGA - DFT | 1.85 | 3.30 | 8.16 |
| | | | vdW-DFT | 1.86 | 2.94 | 8.58 |
| | AB (NN) | | GGA - DFT | 1.85 | 4.54 | 8.14 |
| | | | vdW-DFT | 1.85 | 3.46 | 8.46 |
| | AB (GaGa) | | GGA - DFT | 1.85 | 4.33 | 8.17 |
| | | | vdW-DFT | 1.85 | 3.38 | 8.48 |
| Trilayer (hexagonal type) | AAA | | GGA - DFT | 1.85 | 4.66 | 8.14 |
| | | | vdW-DFT | 1.85 | 3.64 | 8.48 |
| | AA'A | | GGA - DFT | 1.89 | 2.45 | 8.30 |
| | | | vdW-DFT | 1.89 | 2.49 | 8.78 |
| | A'AA | | GGA - DFT | 1.87 | 2.49, 4.46* | 8.20 |
| | | | vdW-DFT | 1.87 | 2.54, 3.59 | 8.62 |
| Tilayer (Bernal type) | ABA (NGaN) | | GGA - DFT | 1.86 | 3.32 | 8.16 |
| | | | vdW-DFT | 1.86 | 2.94 | 8.66 |
| Tilayer (rhombohedral) | ABC | | GGA - DFT | 1.86 | 3.28 | 8.16 |
| | | | vdW-DFT | 1.86 | 2.91 | 8.66 |

* If two numbers are given in one cell, they correspond to the interlayer distance between the bottom two layers and between the top two layers respectively.

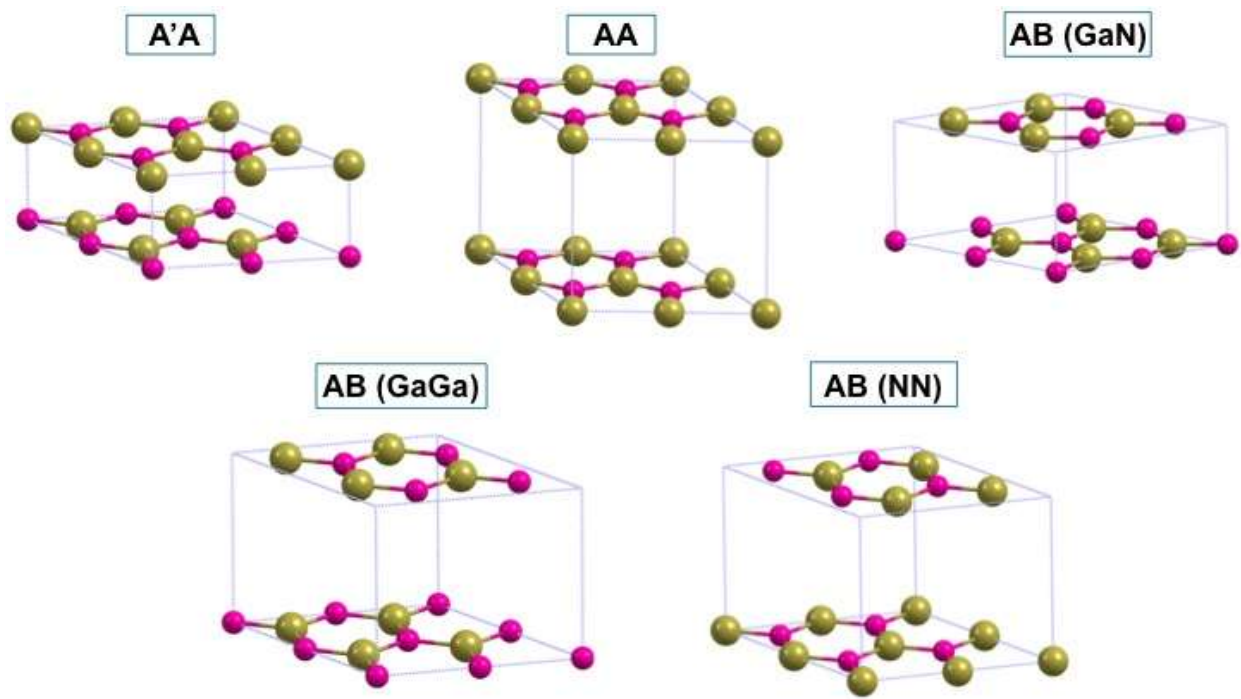

Figure 1   Xu *et al.*

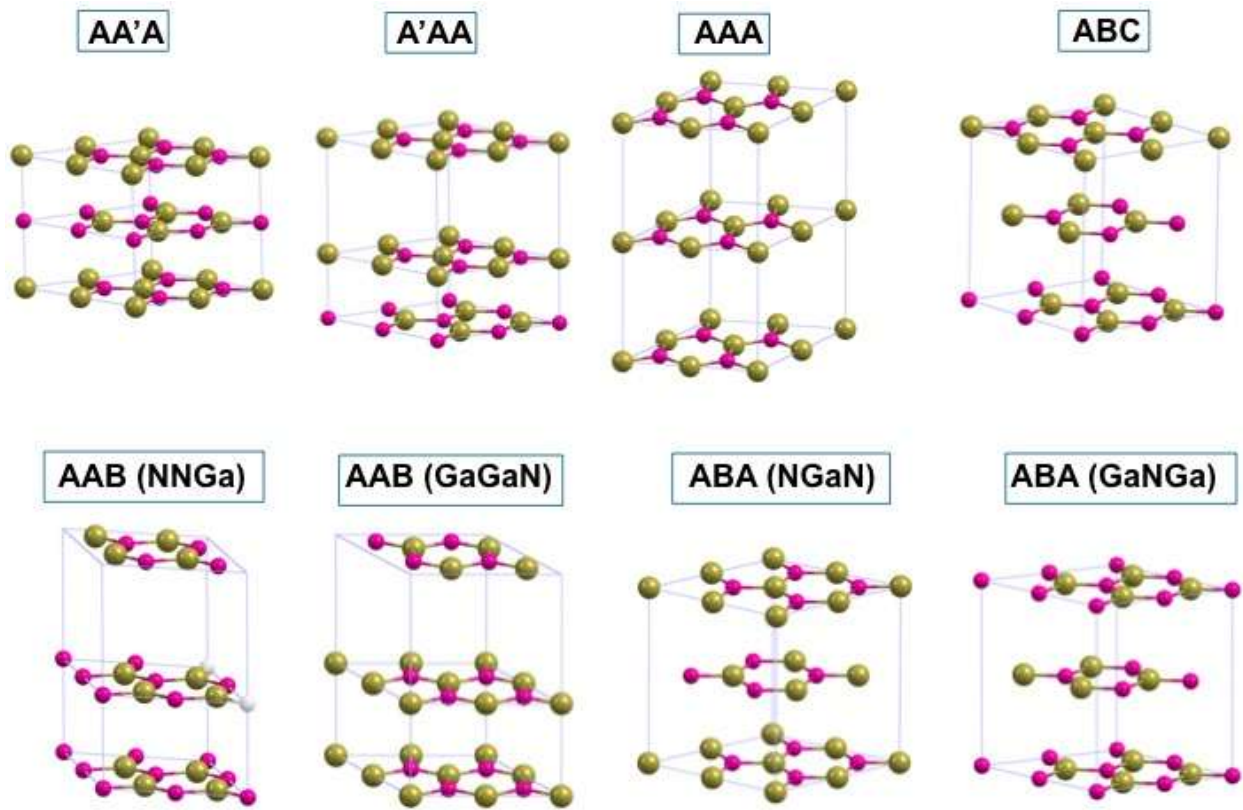

Figure 2   Xu *et al.*

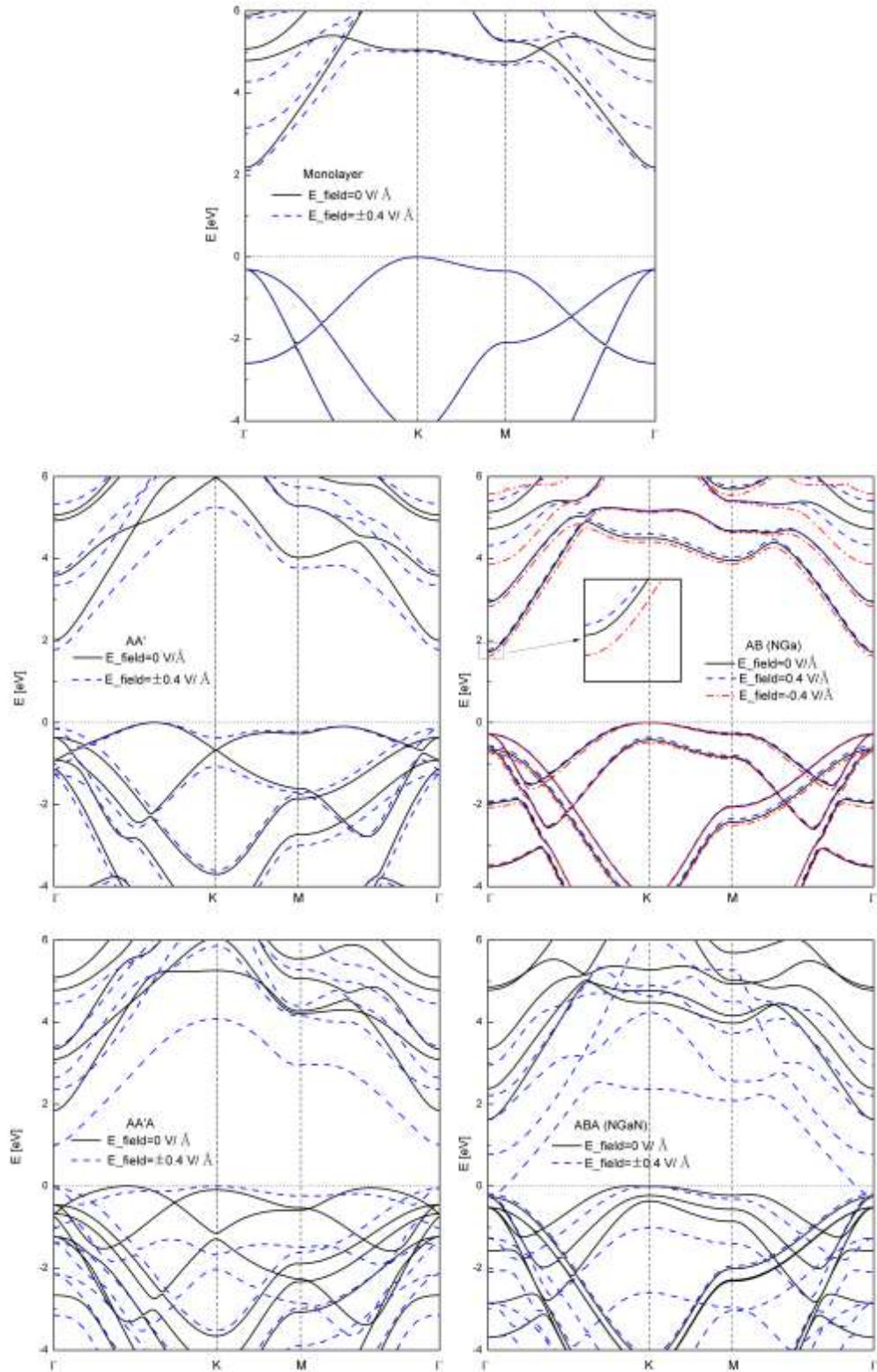

Figure 3 Xu *et al.*

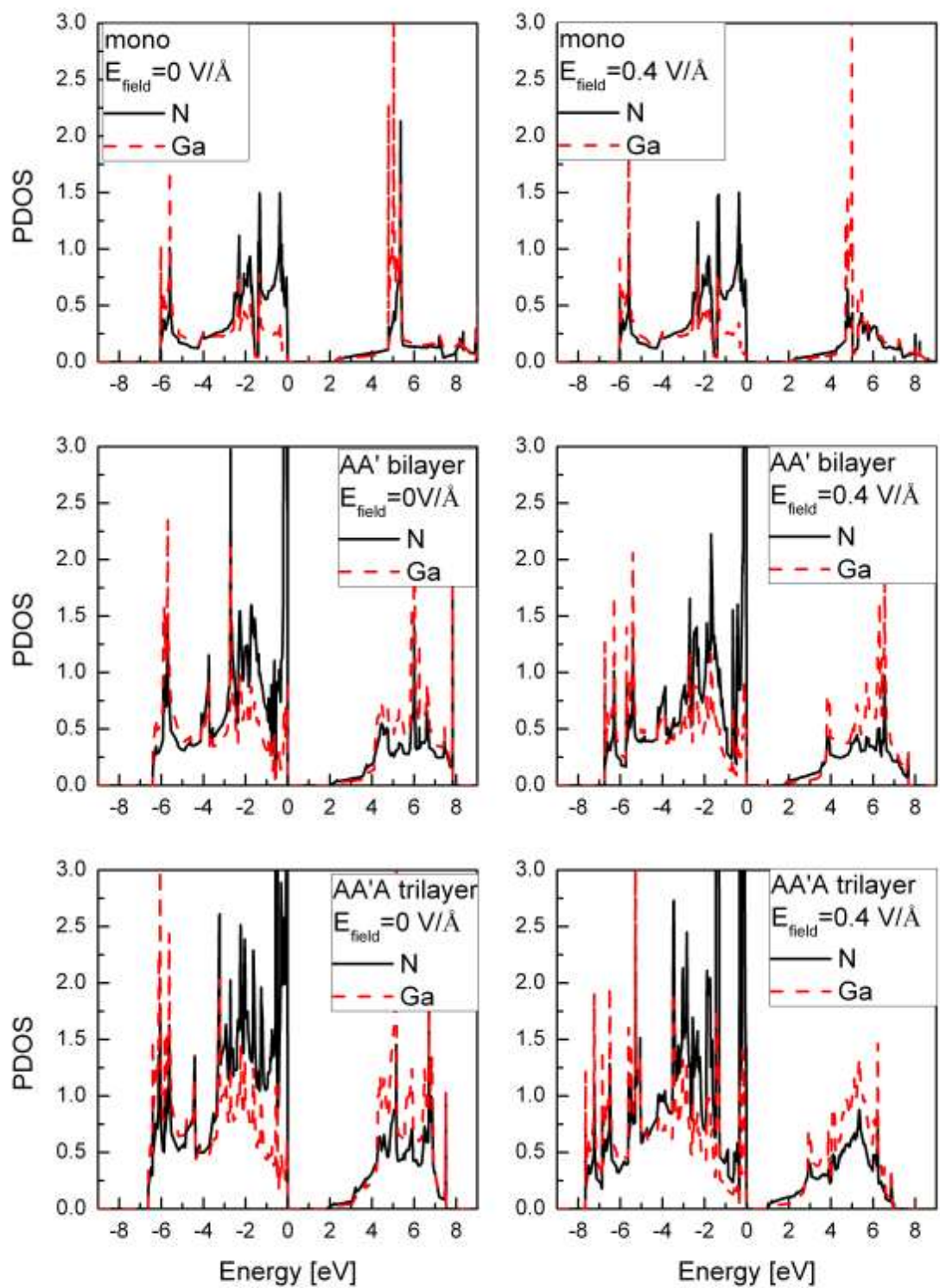

Figure 4   Xu *et al.*

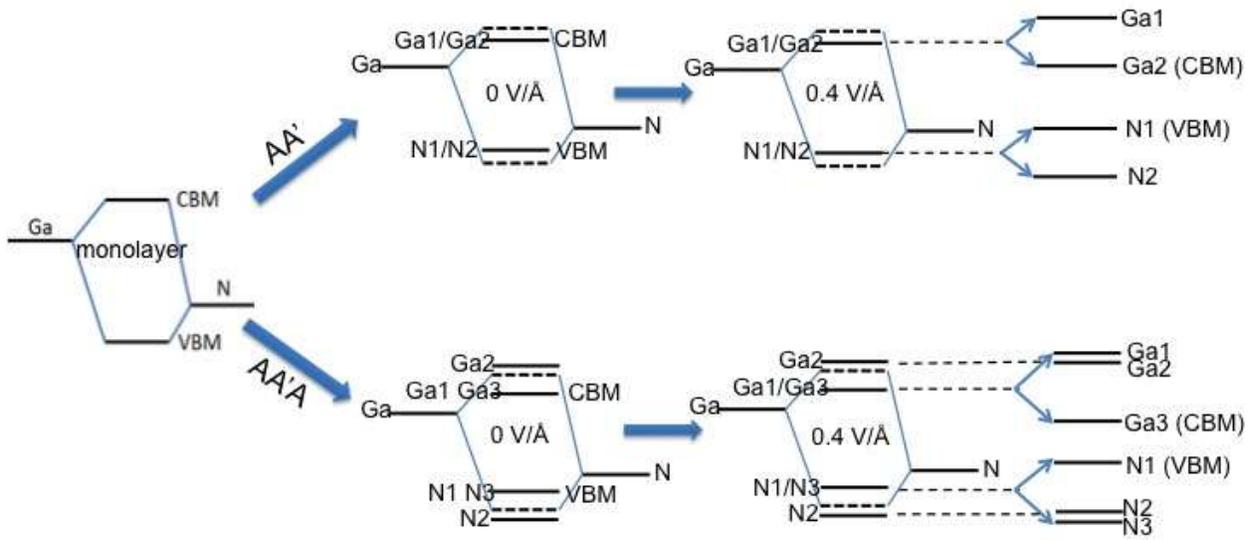

Figure 5   Xu *et al.*

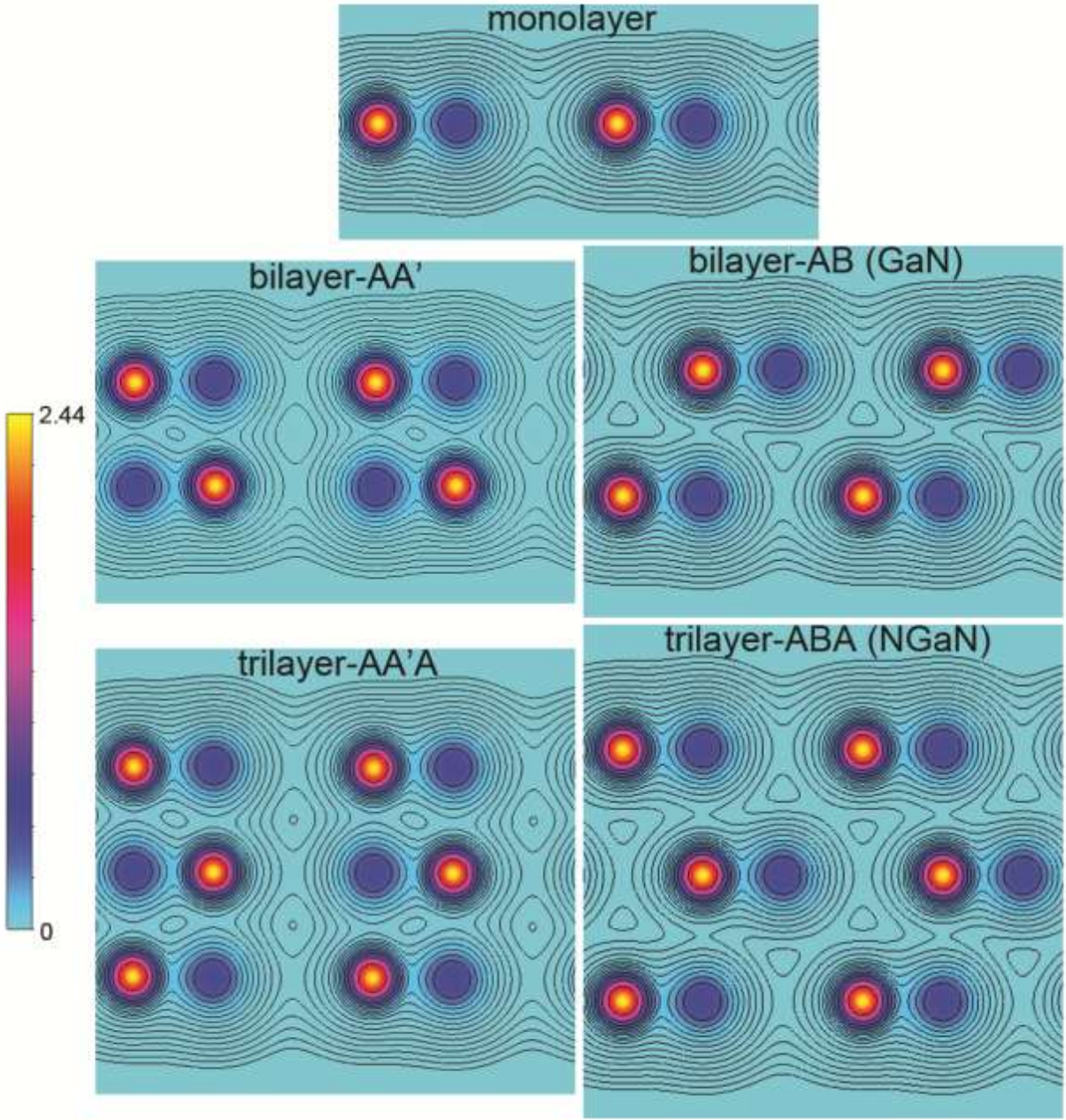

Figure 6   Xu *et al.*

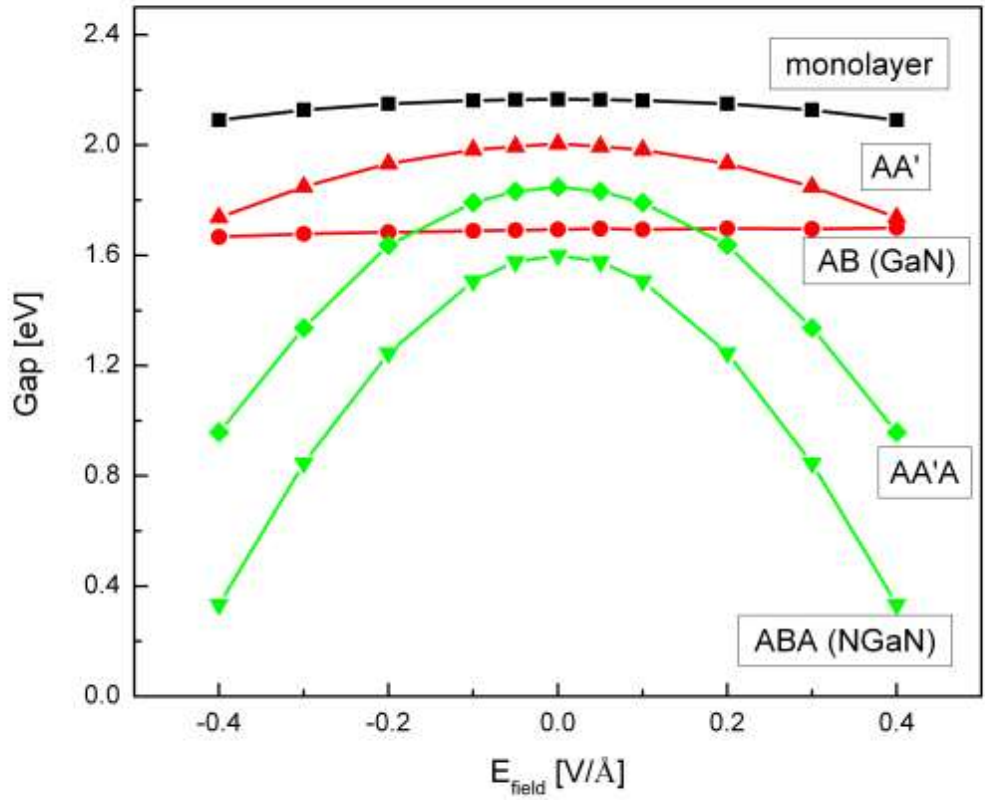

Figure 7   Xu *et al.*